# Challenges & Solutions for above 6 GHz Radio Access Network Integration for Future Mobile Communication Systems

Marcin Rybakowski[1], Krystian Safjan[1], Venkatkumar Venkatasubramanian[1], Arnesh Vijay[1], Laurent Dussopt[2], Ali Zaidi[3], Michael Peter[4], Jian Luo[5], Maria Fresia[6], Mehrdad Shariat[7]

[1]Nokia Networks, [2]CEA-LETI, [3]Ericsson AB, [4]Fraunhofer Heinrich Hertz Institute, [5]Huawei Technologies Duesseldorf GmbH, [6]Intel Deutschland GmbH, [7]Samsung Electronics R&D Institute UK

*Abstract*— **Mobile communication technology has been rapidly evolving ever since its first introduction in the late 1980s. The development witnessed is not just in the refinement of the radio access techniques, but also in the progression towards offering sophisticated features and services to the mobile phone users. To fulfill this ever-growing user demand and market trends, frequency ranges in millimeter wave bands are envisioned for wireless radio transmission. To respond to this trends, the EU-funded mmMAGIC project has been launched and its main objective is to design and develop radio access techniques operating in 6 – 100 GHz bands. When it comes to developing technologies for systems operating these frequency ranges, a major challenge encountered will be in terms of its radio access network integration. Unquestionably, issues at various aspects of physical layer design, channel modelling, architecture, network functions and deployment will be encountered; problems in multi-node and multi-antenna transceiver designs will surface as well. The work carried in this project will address those challenges and propose solutions; but additionally, measure its efficiency against the project specific KPIs set to meet the requirements of the operational future 5G systems. The main intention of this paper is to outline some of the challenges, more specifically to highlight the network integration challenges, and discuss some of its technical solutions. The primary purpose here is to focus towards integrated 5G technology, thereby opening further research avenues for the exploration of new and alternate frequency bands in the electromagnetic spectrum.**

## I. INTRODUCTION

The next generation mobile telecommunication system is predicted to present momentous enhancements, to sustain the growing rise in the traffic volume and the enormous number of use cases [1]. In order to satisfy the requirements for extreme mobile broadband services at high data rate and low latencies, it is mandatory to design future technologies with larger bandwidth than those of its previous generation systems. One potential contender for this would be the frequency bands in 6–100 GHz range; whose potential for use and possibility for deployment are investigated in mmMAGIC (Millimeter-Wave Based Mobile Radio Access Network for Fifth Generation Integrated Communications) [2]. To briefly introduce, the mmMAGIC project was launched in July 2015, with the main objective to design and develop a new system; operating in 5G multi-RAT (Radio Access Technology) context – targeting ultra-dense deployments and ultra-high capacity mobile data services. The project aim, structure and key research areas have been described in previous publications [3], [4]; in which the project intention to offer standard's ready concept for systems operating in 6 to 100 GHz frequency bands has already been stated. A highly pertinent topic discussed here, which is bound to benefit future telecommunications, is the investigation on the issues encountered in network integration of 5G systems (considered as "system of the systems"). In addition, the unique contribution of this publication when compared to previous publications will be in terms of the technical solutions offered in integrating lower RAN layers, previous studies conducted on issues related to core network level integration.

The following paper consists of two major sections with a number of sub-divisions. Section II is the core segment of this paper discussing several topics involving radio access network integration. The section begins in explaining the air interface challenges in terms of inter and intra RAT integration, outlining some basic problems encountered in multi-connectivity and multi-service applications; and examines some of its solutions. Furthermore, the section continues in examining the multi-node and multi-transceiver design challenges, also covering issues related to channel modeling and presents some of its solutions. The last topic covered, will be architectural issues, whereby issues encountered in multi-connectivity and self-backhauling will be addressed. Finally, the paper concludes in analyzing various technical solutions proposed, with a statement of conclusion on its importance for network integration for future 5G systems.

## II. CHALLENGES & SOLUTIONS FOR NETWORK INTEGRATION

### A. Air interface aspects

While having in mind the whole range of frequencies 6 – 100 GHz we first look closer at mm-waves characteristics to formulate the challenges and solutions related to integrate mm-wave systems into 5G network. At mm-wave frequencies, the propagation channel is characterized by high isotropic path loss and penetration loss (causing wave blockage in certain cases), and possible sparsity. High antenna gain is required for transmission, resulting in directional links. In the case of mobile users, strong Doppler effect and intermittent link quality are expected. Moreover, power amplifiers (PA) become less efficient and RF impairments such as phase noise, PA nonlinearity, sampling- and carrier frequency offsets, sampling jitter of Analog to Digital Converters, can be very harsh in mm-wave frequencies.

Any air interface designed for bands in range 6-100 GHz requires proper design, which addresses the above-mentioned challenges and exploits the large contiguous bandwidths, massive number of antennas and dense heterogeneous deployments. The integration of mm-wave nodes among each other or with other air interfaces can help overcome mm-wave specific problems. In example: spotty coverage in mm-wave suggest that users should anyway be served in multi-connectivity between a lower frequency coverage layer and mm-wave. Another example might be issues in mm-wave control signaling which suggest that some signaling (e.g. Radio Resource Control) could be handled via lower frequencies etc. To wrap-up: in order to overcome mm-wave specific problems, the air interface for the systems in 6 – 100 GHz needs to support multi-point connectivity (MC), multiple services, multi-hop backhaul, and potentially also device-to-device (D2D) communication including vehicle-to-vehicle (V2V) as illustrated in Fig. 1.

First, multi-connectivity refers to the case that a user is connected to more than one network node simultaneously. The network nodes can operate at the same frequency or different ones, e.g. sub-6 GHz and above 6 GHz. On the one hand, multi-connectivity is necessary for integrating mm-wave network to the overall 5G network; On the other hand, multi-connectivity is a key to mitigate intermittent link quality due to movement and frequent handover, outage due to blockage etc. Multi-point connectivity can be implemented in different ways -- there can be joint or coordinated transmissions from multiple access nodes that are operating on different frequencies with possibly different types of information being carried through different access nodes. The air interface design including frame structure and numerology of mm-wave RAT has to allow uplink (UL) - downlink (DL) separation (i.e., uplink and downlink transmissions via different nodes), provision of control information using one access node (e.g., macro site) and user data using another access node (e.g., small cell), and to allow access nodes with larger coverage area (e.g., macro cell) to broadcast system information and the densely deployed access nodes with small coverage area to only transmit user specific signals. Furthermore, initial access schemes need to be developed which can efficiently identify potential links to multiple access nodes and make proper selection. Finally, to support multi-connectivity across different frequencies, the air interface of different frequencies should have certain harmonization, e.g. alignment of the frame structures in time, numerologies based on the same reference clock etc.

Second, providing multi-service is foreseen to be a key requirement of 5G and should be supported by mm-wave RAT. Multi-service implies coexistence of different types of UE's, e.g. high-end and low-end, as well as the support of multiple service classes by the RAT. Correspondingly, diverse levels of RF impairments of different UE categories need to be managed, which is challenging. Multi-service also implies different sets of KPI requirements, e.g. data rate, latency, reliability and energy efficiency. Thus, flexibility of air interface is required to support such different requirements, especially some extreme cases that require high data rate, high reliability and low latency simultaneous. Moreover, such requirements may vary in DL and UL differently with different services. A subset of 5G service has to support high mobility, e.g. moving hot spot, which poses challenges in managing strong Doppler effect and dynamic blockage. To address the above challenges, the waveform, frame structure and transceiver design have to be robust against hardware impairments. Transceiver schemes, e.g. equalization, should take into account joint compensation of impairments depending on the expected impairment level. The radio-interface should allow multiple operation modes, e.g. with different waveforms, to address different requirements of multiple services. Further, it is important to have a frame structure that allows flexible adaptation to different DL and UL data rates. The TDD frame structures with flexible DL/UL switching time are promising candidates. To address mobility, transceiver schemes can exploit channel sparsity to combat Doppler effect. Furthermore, multiple links (multi-connectivity) should be exploited to minimize fluctuation of link quality (in case of dynamic blockage). In the case of link interruption, fast link recovery mechanism is necessary.

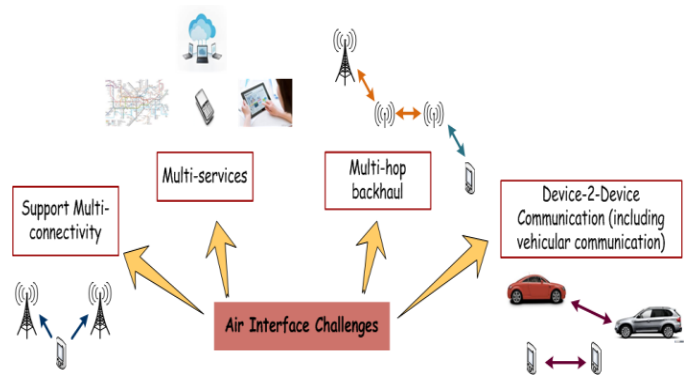

Fig. 1. Main areas of challenges for air interface design for integrated RATs.

Third, multi-hop backhaul including self-backhaul is an important way to realize dense deployment under reasonable cost. The large available bandwidth of mm-wave is a key enabler for self-backhaul, i.e. that the operating both backhaul and access jointly with mm-wave air interface. Usually, backhaul and access links have different channel characteristics, e.g. the backhaul channel is more static and higher probability for LOS, and different requirements. In the above mentioned joint or coordinated transmissions from multiple access nodes for enhancing data rates and providing reliable connection, strict delay requirements are imposed on backhaul links. Thus, accommodating both backhaul and access under one air interface is challenging. Similar to the support of multi-service, different operation modes can be designed for backhaul and access. Further, the frame structure and multiple access scheme should be designed properly so that the DL and UL transmissions of backhaul and access links can be multiplexed efficiently. To address the strict delay requirements, efficient re-transmission schemes are necessary, especially in multi-hop backhaul.

Fourth, D2D communication, including V2V are important feature of 5G and should be potentially supported by mm-wave RAT. While some D2D targets high data rate, others target low latency, e.g. critical machine type communication. One challenge for operating D2D/V2V at higher mm-wave frequencies is the device mobility, especially dual mobility (both devices are moving). A specific problem is the beams alignment. Under dual mobility, the beams alignment has to be done in a short time slot and may need to be repeated when relative positions of both devices change considerably. A further challenge is link blockage, which could further exacerbate when the vehicular density is high. To address dual mobility, fast initial beams alignment and re-alignment mechanisms have to be developed. To address blockage problem, multi-hop D2D connections can be exploited. To fulfill latency, reliability and availability requirements the amount of resources for transmission is crucial. Interestingly, blockage can even be exploited to further increase the reuse factor, since blockage can also block interference signal. Considering also highly directional links, multiple D2D links can aggressively reuse radio resources available at mm-wave frequencies. Finally, by exploiting the large bandwidth, distributed resource allocation can be applied with low probability of collisions.

*B. Multi-node, multi antenna transceiver design aspects*

Enabling mm-wave radio access will require large and flexible antenna array technologies to be developed with capabilities beyond beamforming mode according to e.g., propagation conditions, services etc. These multi-antenna and multimode transceiver solutions shall enable flexible support for a multitude of services and use cases, number of users/beams, different connectivity distances, edge-less user experience influenced by mobility and severe signal blockage and coverage holes with high energy and cost efficiency.

The candidate multi-antenna transceiver architectures can be classified into four main architectural families, which are briefly described in the following (Fig. 2):

a) No beamforming (or fixed-beam) systems are the simplest architecture with a single baseband (BB) and RF antenna. This is the architecture that is employed in current mobile networks at sub-6 GHz frequencies for terminal and base-stations operating in single-input single-output (SISO) mode, as well as for point-to-point links (e.g. backhauling) at mm-wave frequencies.

b) Analogue beamforming architectures are used today in high-end mm-wave systems such as radars and in recently demonstrated 60 GHz indoor short-range communication systems (WiGiG, IEEE 802.11ad). Beamforming functions can be implemented in continuous or discrete (switched) form depending on requirements.

c) Digital beamforming architectures provide the best performance in terms of data rates and multiplexing capabilities due to having the highest level of flexibility; they are used today in MIMO systems at sub-6 GHz frequencies (i.e. IEEE 802.11ac) with up to 4 antennas in commercial products, but not at mm-wave frequencies.

d) Hybrid beamforming architectures mix digital and analogue beamforming techniques in order to provide a performance trade-off on flexibility, multiplexing, power consumption and cost.

Each of these architectural families offer distinct capabilities based on different requirements. Fixed-beam and analogue beamforming systems have been demonstrated nearly in every band of the mm-wave spectrum at least at prototype level. Many commercial solutions for fixed-beam point-to-point communications (backhaul) are available in Ka (26–40 GHz), V (57–66 GHz), and E (71–86 GHz) bands. In the 60 GHz band, several products with analogue beamforming have been commercialized as well. In contrast, digital or hybrid beamforming architectures have hardly been demonstrated to date at mm-wave bands, to the best of our knowledge. First experimental results of hybrid beamforming systems have been presented recently [7], [8], [9]. On the other hand, fully digital large-scale antenna systems at mm-wave frequencies appear today as a challenge in terms of power consumption, integration and cost, especially since a large number (several tens or hundreds) of antennas is expected to be required for mm-wave access with ranges up to 100–200 meters in outdoor conditions.

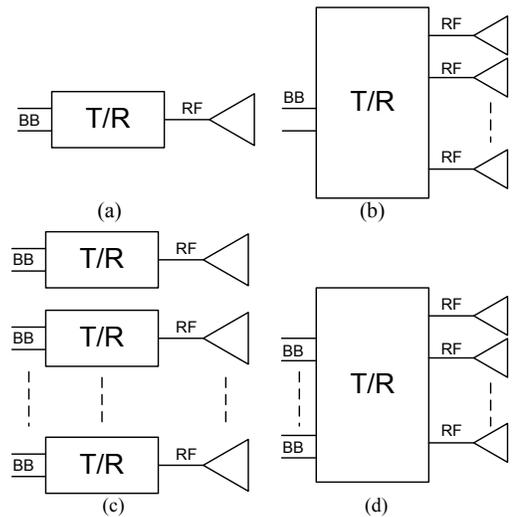

Fig. 2. Schematic representation of the transceiver architectures: (a) no beamforming, (b) analogue beamforming, (c) digital beamforming, (d) hybrid beamforming.

*C. Channel modelling*

In the last decade, significant effort has been made to elaborate and continuously improve channel models for cellular networks operating in the traditional frequency bands

[1], [6], [7]. However, as prerequisite for link- and system-level simulations of future 5G systems, accurate channel models above 6 GHz are required. The ultimate goal of related activities would be to provide one single comprehensive model, which is valid over the entire frequency range from 6 to 100 GHz. However, this objective is exceptionally challenging.

At mm-wave frequencies, the propagation characteristics significantly differ from the characteristics at lower frequencies. With increasing frequency, radio propagation more and more approaches the behavior of optical propagation and increasingly relies on the line of sight (LOS) path or strong reflections rather than on diffuse components [8]. Therefore, in addition to the features of state-of-the-art models, future models need to properly incorporate shadowing/blockage effects as well as time variance caused by a dynamic environment. Blockage can be induced by trees and street furniture, traffic or walking or people. 3D spatial capabilities become crucial in order to support the simulation of beamforming and MIMO techniques, which are essential for mm-wave access over distances corresponding to small cell size (tens to hundreds of meters). A hybrid model involving a stochastic as well as a deterministic component may be needed to accurately incorporate these effects [8].

Deployment of 5G systems is expected to be very diverse and comprises e.g. outdoor macro cells, providing coverage in street canyons, down to small cells in stadiums, shopping malls and offices. A comprehensive model should not only cover the various environments, related antenna heights and outdoor-to-indoor propagation, but even be capable to reproduce inter-frequency and inter-site correlations to enable the simulation of heterogeneous access at multiple frequencies. Moreover, the model must support up to several GHz of channel bandwidth. The overview of the directions of extending the state-of-the-art channel models to be used with 5G systems is shown in Fig. 3.

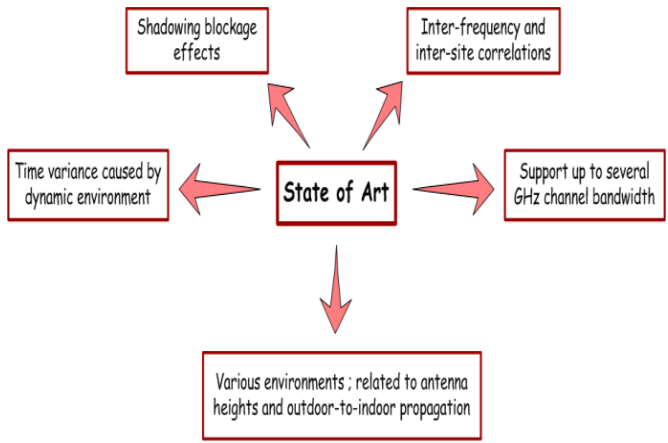

Fig. 3. Extension directions of the state-of-the-art channel models to be used for 5G systems modelling

The elaboration of a channel model is inextricably linked with channel measurements. To provide a solid basis for a frequency-agile model, measurement data for multiple frequency sample points between 6 and 100 GHz must be available for all relevant environments and propagation conditions. They need to comprise a large sample size to allow statistical evaluation and provide temporal as well as directional information, both with high resolution. However, this is extremely challenging, since suitable multi-antenna hardware is not available yet and the development of dedicated solutions for multiple frequency bands up to 100 GHz is hardly possible at reasonable expense. Measurement setups with mechanically steerable antennas yield directional information, but it is hard to achieve a large sample size due to the long measurement duration for each point in space. Channel sounding with omnidirectional antennas is beneficial to derive path loss models, capture all multipaths and also time variance – provided that the system can measure sufficiently fast. However, directional information is not provided directly.

In mmMAGIC, seven partners will pool their capabilities and conduct around twenty measurement campaigns in more than eight frequency bands from 6 to 100 GHz, explicitly including the edge frequencies. The campaigns involve channel sounder setups with multi-frequency (up to four bands in parallel) and ultra-wideband capabilities (up to 4 GHz bandwidth). The measurement campaigns will be complemented by map-based simulations based on ray tracing and point cloud models. This approach enables to obtain additional information about the channel while relying in the solid basis of measurements.

A white paper has been released, which briefly describes the measurement and modelling plans [9]. The deliverable D2.1 [10] summarizes early available measurement results and the mmMAGIC initial channel model. Upcoming data will be used to update the parameter tables and further extend and refine the model towards the mmMAGIC final channel model.

### D. Architecture aspects of integration, logical and physical deployment, RAN functions.

Considering integration aspects of RAN architecture shouldn't be done in separation to envisioned deployments – which determines availability of the information in both physical locations and logical entities of the network and indicates potential for information exchange taking into account transportation constrains. The main novelty in the integration approach is to design integration solutions at the lowest possible level in the protocol stack. In this subsection, while studying challenges and solutions for architectural aspects of integration we look into four key RAN functions supporting integration: user and control plane split, fast scheduling and fast handovers, multi-connectivity and self-backhauling.

As it was mentioned above it is beneficial to consider 5G network architecture jointly with envisioned deployments. In mmMAGIC, we consider three deployment variants for above 6 GHz systems that will be used to investigate and evaluate integration solutions – the operation variants are: standalone operation of above 6 GHz RAT, non-standalone operation of the above 6 GHz RATs (i.e. with support from below 6 GHz RAT) and mm-wave as an enabler for other technology operation as shown in Fig. 4. [1].

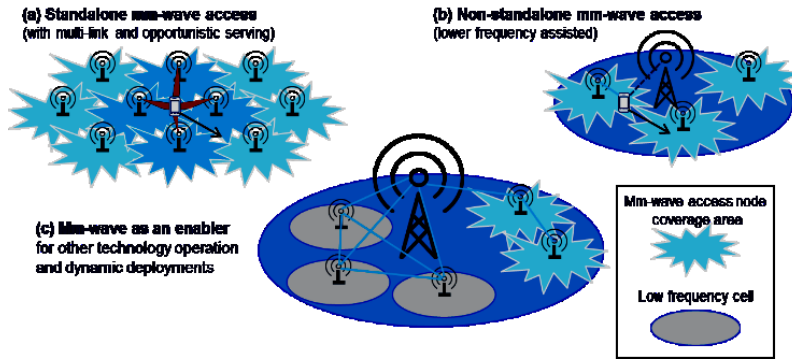

Fig. 4. Illustration of possible deployment of mm-wave technologies: (a). standalone operation;

In case of network integration and architecture the non-standalone deployment is challenging - especially when taking into account mobility and coverage aspects. The non-standalone deployment will utilize joint deployment of nodes operating above 6 GHz with nodes operating at lower frequency bands – below 6 GHz such as LTE-A or new 5G wide area RAT. The physical realization of the APs can be implemented in the form of multi-RAT Access Points (AP) or neighboring APs of different RATs.

Considering the currently deployed 3G and 4G systems, an inter-RAT integration is feasible only at the higher network layers and core network involvement for radio resource control tasks, such as reselections and handover processes. However, the 5G system requirements for an edge-less user experience can lead to the realization of inter-RAT integration at the lower network layers as well. The reason being: some critical control mechanisms at the physical layer, like the delivery of scheduling information, demand strong reliability; and therefore require tight integration with an umbrella layer at lower frequencies. The target here will be the integration of RATs specifically covering load balancing amongst different systems and the selection of optimum RAT for each service. Specifically in case of non-standalone scenarios additional benefits can be expected, if the above 6 GHz systems are supported by technologies operating below 6 GHz. Network configuration information and parts of control plane information can be exchanged on lower frequency bands e.g. to announce mm-wave systems present within coverage area, and to leverage initial access or handover procedures. The potential solution is that mm-wave systems will have user and control plane splitting, where the low frequency bands can support mm-wave systems in improving the reliability of the control plane. The considered principle that helps to make right selection of the RAT is to prefer sending the control information over licensed bands rather than un-licensed or lightly-licensed ones. Due to the significantly better coverage at lower frequency bands, the control and setup of the mm-wave link is more reliable; hence, a part of the control communication is transmitted on the lower frequency band. Another solution targeting provision of high reliability of control information delivery is RRC (Radio Resource Control) diversity – what refers to the situation when UE is controlled by multiple RRC instances (e.g. located in neighboring APs).

In order to serve users with moderate and high mobility, the networks need to have fast scheduling and fast handover mechanism. The design of the beam-forming/-steering may impact the scheduling/handover process. In outdoor scenario with moderate and high speed moving users, non-standalone operation may be dominantly used compared to standalone operation, since the lower frequency may provide more robustness to blockages and hence may reduce the frequency of handover. Further, the non-standalone RATs should also allow for fast seamless and reliable mobility and aggregation handling between the RATs, with efficient management and pooling of resources for optimum performance by help of the lower frequency band coverage layer. The initial access could be enhanced by allowing first step of system access through RAT on lower frequency band and later adding the second access step through RAT on above 6 GHz frequency bands.

The multi-connectivity was already discussed in this paper in section II.A from the physical layer perspective. In this section RAN function perspective is used to analyze how this solution can support integration.

Multi-connectivity, is an enabling concept extending carrier aggregation (CA) and Coordinated Multipoint (CoMP) for inter-eNodeB and inter-RAT cases. To support MC, the air interface should consider a diverse set of requirements on backhaul, from ideal dedicated ones with sufficient bandwidth, low delay and low jitter to more relaxed backhaul with higher levels of impairments.

The multi-connectivity plays a key role to provide edge-less user experience and to achieve broadband access everywhere (in particular in dense areas and targeted hot spots). In terms of mobility, efficient transport layer is required to manage potential frequent inter-RAT handovers. Particularly, for cloud services and moving hotspots, higher target mobility introduces faster handover / beam forming requirements due to possible blockages. This needs integration of low and high frequency networks on lowest possible level in order to mitigate spotty coverage of mm-wave APs. MC can be instrumental in these cases to control a group (cluster) of mm-wave cells and to ensure smooth operation. Supporting MC is more challenging for standalone mm-wave RAT. This requires intelligent and dynamic AP clustering algorithms taking into account the required time for the MC initiation as well as the coverage per mm-wave cluster to ensure edge-less user experience. In this direction, context-learning algorithms can be exploited to have better clustering procedures in presence of MC. From RRM perspective, more flexible resource management between high and low frequency band is required with MC considering different KPIs including energy and resource consumption. This may need revisiting the protocol stack to modify and introduce new procedures and functionalities. In particular, the cost of supporting a specific rate may not be the same in different frequency bands (RATs) in terms of energy consumption, level of resulting inter-cell interference and required signaling overhead. Therefore, resource management function needs to consider all these factors in a multi-band resource scheduling.

Wireless self-backhauling is a promising solution to support emerging networks via autonomously establishing backhaul connectivity to existing network structures, in particular, where dedicated backhauls become cost-prohibitive and difficult to deploy. In self-backhauling, backhaul and access links share the same radio resources. Moreover, the backhaul link may be shared among several APs. Therefore, dimensioning of radio resources is one of the fundamental aspects to address. Another fundamental design aspect is the role of self-backhauled APs under the coverage of a donor AP (i.e. Amplify and Forward, Decode and Forward, Full L3 functionality). Number of hops is another important front, affecting many technical aspects of a self-backhauled network. Increasing the number of hops provides greater deployment flexibility but adds on to the latency. When considering protocol stack design for self-backhauling the two foreseen most challenging aspects are support for rapid rerouting and multiple hop links. The overall protocol stack needs to be designed in the way that ensures seamless operation when UEs move between APs and fixed cells. Mobility introduces additional complications into self-backhauling (e.g. moving hotspots) and requires enhanced level of dynamics for in-band backhaul networks. Therefore, dynamic bandwidth partitioning algorithms between radio access and backhaul should be adopted.

## III. CONCLUSIONS

In this paper the various challenges that has to be faced in order to adopt mm-wave technology in future mobile systems have been described. In particular the key aspects of the network integration have been analyzed from different directions.

Problems and possible solutions to design air interfaces for supporting network integration have been studied. In particular, amongst all the problems to be addressed when designing the air interface, four of them (named multi-point connectivity, multiple services, multi-hop backhaul, and device-to-device communication) have been analyzed, and possible solutions to be studied by the mmMAGIC project have been sketched.

The channel characteristics and the open points that still need to be addressed in order to elaborate a good channel model for frequency above 6 GHz have been also described. A successful design of the air interface, in fact, cannot be disconnected from a correct channel modelling. The measurement campaign conducted and how the channel modelling problem will be attacked by the mmMAGIC project have been discussed.

The potentiality of non-standalone deployments, and possible solutions to exploit the advantages of inter-RAT of integration have been discussed. Efficient solutions, studied by mmMAGIC project, in terms of C/U plane split for edgeless user experience, robustness to blockage when users are moving and for optimizing the self-backhauling have been pictured.

ACKNOWLEDGMENT

The research leading to these results received funding from the European Commission H2020 programme under grant agreement n°671650 (mmMAGIC project).

REFERENCES

[1] NGMN Alliance, 'NGMN White Paper' Feb. 2015 (available online www.ngmn.org/uploads/media/NGMN_5G_White_Paper_V1_0.pdf)
[2] mmMAGIC project. Official website: https://5gmmmagic.eu/
[3] Maziar Nekovee, Peter von Wrycza, Maria Fresia, Michael Peter, Jacek Gora, Jian Luo, Milos Tesanovic, "Millimetre-Wave Based Mobile Radio Access Network for Fifth Generation Integrated Communications (mmMAGIC)", EuCNC, July 2015
[4] "Use cases, KPIs and spectrum considerations for 5G system in the 6-100GHz range" NGMN Forum in Montreal, October 2015
[5] 3GPP TR 25.996 V10.0.0, "Spatial channel model for multiple input multiple output (MIMO) simulations," 2011.
[6] WINNER II deliverable D1.1.2 V1.1, WINNER Channel Models, 2007. Online: http://www.ist-winner.org/deliverables.html.
[7] L. Liu, C. Oestges, J. Poutanen, K. Haneda, P. Vainikainen, F. Quitin, F. Tufvesson, and P. Doncker, "The COST 2100 MIMO channel model," Wireless Communications, IEEE, vol. 19, no. 6, pp. 92–99, Dec. 2012.
[8] A. Maltsev, A. Pudeyev, I. Karls, I. Bolotin, G. Morozov, R. Weiler, M. Peter, and W. Keusgen, "Quasi-deterministic approach to mmmm-wave channel Modeling in a non-stationary environment, IEEE Globecom," Workshop on Emerging Technologies for 5G Wireless Cellular Networks, Austin, TX, USA, Dec. 2014.
[9] ICT-671650 mmMAGIC White Paper 2.1, "6–100 GHz channel modelling for 5G: measurement and modelling plans in mmMAGIC," Feb. 2016. Online: https://5g-mmmagic.eu/results/.
[10] ICT-671650 mmMAGIC, Deliverable 2.1, "Measurement campaigns and initial channel models for preferred suitable frequency ranges," Mar. 2016. Online: https://5g-mmmagic.eu/results/.
[11] X. Gu, et al., "W-band scalable phased arrays for imaging and communications," IEEE Comm. Mag., Apr. 2015.
[12] A. Maltsev et al., "Performance evaluation of the mm-wave Small Cells communication system in MU-MIMO", EuCNC 2015.
[13] W. Roh, et al., "Millimeter-wave technology beamforming as an enabling technology for 5G cellular communications: theoretical feasibility and prototype results," IEEE Comm. Mag., Feb. 2014.